# Epitaxial integration of the intrinsic ferromagnetic semiconductor GdN with silicon technology

F. Natali[1*], N. O. V. Plank[1], B. J. Ruck[1], H. J. Trodahl[1], F. Semond[2], S. Sorieul[3] and L. Hirsch[4]

[1]MacDiarmid Institute for Advanced Materials and Nanotechnology, School of Chemical and Physical Sciences, Victoria University of Wellington, PO Box 600, Wellington 6140, New Zealand

[2]Centre de Recherche sur l'Hetero-Epitaxie et ses Applications, Centre National de la Recherche Scientifique, Rue Bernard Gregory, Sophia Antipolis, 06560 Valbonne, France

[3]Centre d'Etudes Nucléaires de Bordeaux-Gradignan IN2P3, UMR 5797, Université de Bordeaux 1, Chemin du solarium BP120, 33175 Gradignan Cedex, France

[4]Université Bordeaux 1, Laboratoire d'Intégration du Matériau au Système – CNRS UMR 5218. ENSCPB, 16, Avenue Pey Berland, 33607 Pessac Cedex, France

[*]Corresponding contact: franck.natali@vuw.ac.nz

A major challenge for the next generation of spintronics devices is the implementation of ferromagnetic-semiconductor thin films as spin injectors and detectors[1,2]. Spin-polarised carrier injection cannot be accomplished efficiently from metals[3], and coupled with the rarity of intrinsic ferromagnetic semiconductors this has driven intensive study of diluted magnetic **semiconductors[4]. Chief among these is the doped III-V compound (Ga,Mn)As. These materials suffer from a number of drawbacks; they (i) require magnetic-ion doping well above the solubility limit, and (ii) must be hole doped to above the degenerate limit, preventing independent control of the carrier concentration and charge sign. Here we demonstrate the first epitaxial growth of a recently-characterised *intrinsic ferromagnetic semiconductor*, GdN, on silicon substrates, providing an essential step on the way to integrate new spintronics functionalities into Si-based technology. The films have been characterised as regards their growth toward fully relaxed GdN, the density and mobility of their carriers, and their magnetic behaviour.**

The rare-earth nitrides (RN) were studied already 50 years ago, but only recently have been grown with sufficient stoichiometry to establish their properties confidently. Across the rare-earth series they display a wide range of transport and magnetic properties, though most are ferromagnetic semiconductors at low temperature. GdN is the most thoroughly studied, in part because of the maximum spin moment of 7 $\mu_B$ and zero orbital angular momentum in the half-filled 4$f$ shell of $Gd^{3+}$. Historically its conducing nature was uncertain, due to doping by a high concentration of N vacancies[5,6,7,8]. GdN films with low vacancy densities have resistivities of moderately-doped semiconductors. Its optical band gap is 0.9 eV in the ferromagnetic phase, and the band structure tuned to this gap shows a semiconductor with an indirect gap of 0.43 eV[9]. That band structure has been confirmed by resonant and non-resonant x-ray spectroscopy[10]. The spin splitting is about 0.4 eV (4500 K) in both band edges, with the majority spin (parallel to the Gd 4$f$ shell spin) having the lower energy in the conduction band and the higher energy in the valence band. Thus carriers at both edges, electrons and holes, are in majority-spin bands. The minority-spin edges are unoccupied at ambient temperature. Any device (diode, transistor) that requires doping or accumulating carriers into the band edges will involve transport of carriers with only that majority spin state. In the injection mode the electrons or holes that can be injected into or from another semiconductor will likewise carry only that majority spin.

Eu, with its propensity to form the divalent state, forms also a metastable mono-oxide, EuO, which has already been grown as a half metal on Si[11]. It shows magnetic properties similar to GdN, but with a strongly differing band structure. Thus, for example, the 4$f$ character of its valence band renders it too flat to form usefully mobile holes[12,13].

GdN adopts a face centred cubic (NaCl) structure with a lattice parameter of 0.498 nm. To date epitaxial growth has been carried out onto YSZ(100)[14], MgO(100)[15], and GaN(0001)[16] surfaces. However, silicon remains the most well-developed and heavily-exploited semiconductor, so any future exploitation will place a premium on integration with Si. A similar premium was placed on epitaxial III-V/Si integration, which has recently led to the monolithic integration of GaN-based transistors and silicon MOSFETs on a silicon substrate[17].

The 7.5% lattice mismatch between GdN and Si, though relatively severe, is not an insurmountable problem; more problematic is the reactivity of Gd with Si, encouraging the formation of a metallic GdSi$_x$ layer[18]. Our solution has been to grow GdN on an epitaxial AlN(0001) buffer grown *ex situ* on Si(111) by molecular beam epitaxy[19]. These templates were then placed in an UHV chamber with a base pressure of ~8 × 10$^{-9}$ Torr where the GdN was grown at temperatures of 650-700°C under 3.5 × 10$^{-5}$ Torr of pure N$_2$. N$_2$ reacts with Gd at the surface to form an epitaxial GdN layer, even in the absence of activated N$_2$, as has been demonstrated for polycrystalline RN films[7,20] but this is its first demonstration for epitaxial growth. To prevent decomposition of GdN in air a polycrystalline AlN passivating cap was grown at 300K by evaporating Al in the presence of activated nitrogen; the final structure is illustrated in Figure 1a. The composition, structure and magnetic and transport properties of the epitaxial (111)-oriented GdN films were investigated with Rutherford backscattering spectroscopy (RBS), reflection high-energy electron diffraction (RHEED), X-ray diffraction (XRD), electron transport and magnetisation measurements.

The development of the in-plane lattice parameter during the growth (Figure 1b) shows an onset of plastic relaxation after only ~2.5 monolayers (MLs), as expected for the +13% lattice mismatch between AlN and GdN. Two dimensional growth is observed after the onset of plastic relaxation (Figure 1b inset). Further growth allows a rapid evolution to a fully relaxed GdN layer after only 6MLs. The in-plane lattice parameter deduced from RHEED along the Si[110] azimuth is 3.52 Å, in agreement with the bulk GdN value ($a_{GdN(111)} = a_{GdN}/\sqrt{2} = 3.521$ Å). On increasing the thickness beyond ~25 nm the RHEED pattern becomes spottier, signalling a roughening surface.

Figure 2 displays the XRD 2θ-scan confirming the epitaxial character. In addition to the peaks of AlN and Si we observe only the (111) and (222) reflections of GdN. Clearly the hexagonal face of AlN favours a fully (111)-oriented GdN film. Using the Scherrer formula, we extract a coherence length of ~15 nm, somewhat smaller than the ~26nm for GdN grown on nearly lattice-matched YSZ[14]. This is not unforeseen for the 13% lattice mismatch between AlN and GdN and the high dislocation density in the AlN buffer[21].

RBS measurements were carried out in random and channelling geometries to investigate the epitaxial nature of the GdN layer. Figure 3 shows the random (blue) and aligned (black) spectra of a 20.5 nm thick GdN layer. The reduction in yield associated with channelling is clear evidence of the excellent epitaxial quality of the GdN layer; note that the reduction of the Gd signal is by a very similar fraction to that found from both the Si substrate and the AlN buffer layer. Although the nitrogen RBS signal overlaps with that of silicon, the resolution is sufficient to give the Gd:N ratio of 1.0 ± 5%.

Figure 4 shows the temperature- and field-dependence of the magnetisation measured with a superconducting quantum interference device (SQUID) for an in-plane field orientation; a Curie temperature of $T_c \approx 65$ K is deduced from both the onset of hysteretic magnetisation and by the temperature dependence of the inverse susceptibility in the paramagnetic phase (Fig. 4a). The magnetic hysteresis at 10 K shows a coercive field of 125Oe and a saturation magnetization $M_{sat}$ of 7.5 ±0.5 $\mu_B/Gd^{3+}$, in agreement with the expected 7.0 $\mu_B/Gd^{3+}$ for a half-filled 4$f$ shell (Fig. 4b). The coercive field is a factor of about two smaller than we have reported in films with a similar crystallite size grown on YSZ, suggesting that even smaller coercive fields may be typical of well-ordered GdN[14]. Note that oxygen contamination as small as a few percent is reported to increase the coercive field to thousands of Oe and to decrease the Curie temperature of GdN[22]. The very small coercive field we measure supports the absence of significant oxygen that has also been noted in RBS data, establishing the efficacy of the AlN cap as a passivation layer.

The 65 K Curie temperature of GdN is easily within the range of low-cost cryocoolers that have been developed for exploitation of the cuprate superconductors. Nonetheless it is interesting to consider whether there are opportunities for raising it to at least above the boiling point of liquid nitrogen. There are reports of higher transition temperatures, most notably by the introduction of carbon in the network[23] and furthermore there is both a theoretical and experimental suggestion of $T_C$ enhancement with electron doping[11,23,24]. These need exploration, though in both cases one loses the benefit offered by freedom to dope independently.

Electron transport measurements have been carried out using a van der Pauw configuration. The Hall effect shows n-type conduction with a carrier concentration of ~$1.1 \times 10^{21}$ cm$^{-3}$ at ambient temperature. The resistivity of 0.1 mΩ.cm then gives a mobility of ~15 cm$^2$/V.s, similar to previous data on a PLD-grown epitaxial GdN layer on YSZ[14] and plasma-assisted MBE grown epitaxial GdN layer on GaN[16]. Assuming that each nitrogen vacancy provides three electrons this suggests vacancy densities of order 1%. From these values and assuming a free electron effective mass $m^*/m = 1$, we calculate a mean free path of 11 nm, comparable to the layer thickness and the crystallite radius (15 nm) inferred from XRD spectra. The temperature-dependent resistivity, plotted in Figure 5, shows a negative temperature coefficient (TCR), broken by the well-established anomaly at $T_C$[7,8,14,16,25]. The high carrier concentration is above the degenerate limit, so that the negative TCR in these data signals weak localisation in the degenerate regime, enhanced in these thin films[26].

Summarising the experiments, we have grown epitaxial films of the ferromagnetic semiconductor GdN on Si(111), preventing the problematic silicide formation by the use of an AlN buffer layer. The films have a Curie temperature of 65 K and a saturation magnetization in agreement with previous data on GdN. They are doped by N vacancies. The demonstration of the growth of high quality GdN films on silicon paves the way for the future integration of this intrinsic ferromagnetic semiconductor on silicon that has potential roles in spintronics.

Gazing further into the future, the rare-earth nitrides form a homologous series, with a lattice constant varying smoothly from 0.5013 nm for CeN to 0.4766 nm for LuN, permitting heteroepitaxy across the entire series. With the exception of the fully occupied 4$f$ shell in LuN they carry a spin moment, provoking the ferromagnetic order that has been observed for most of them. Experimental band structure data are so far meagre, despite an intense theoretical interest. Nonetheless the few that we have studied, SmN, DyN and ErN, are all semiconductors[27,28]. The computed band structures also show very similar behaviour to GdN as regards the RE 5$d$ and N 2$p$ states that form the valence and conduction band edges; as expected it is primarily the 4$f$ bands that differ across the series[29]. Thus, for example, the common spin state at the two band edges is consistent across the series. Their magnetic states vary more strongly. In particular SmN shows full ferromagnetic order among the 4$f$ spins, but with almost complete cancellation of the magnetic moment by a 4$f$ orbital contribution[30]; it is the only known near-zero-moment ferromagnetic semiconductor. It is thus free of a significant fringe field to disturb transport in adjacent Si or other nonmagnetic semiconductors, and the small moment couples so weakly to an external field that is has an enormous coercive field. GdN, in comparison, has a coercive field some three orders of magnitude smaller, so these form an ideal hard- and soft-ferromagnetic pair with potential in memory elements.

Finally we note that the thickness of AlN in this study will prevent direct injection of carriers across the interface in either direction, which was required in order to measure the GdN transport data without the confusion of parallel conduction through the Si substrate. However it is certain that successful epitaxy can be performed on much thinner AlN, as reported for the realisation of group III-nitrides-based devices[17,19]. It is of interest to note that the AlN buffer layer not only prevents the problematic silicide formation but can also act as template for the integration of GdN into the group III-nitrides, a technologically important family for the fabrication of optoelectronic devices and high power transistors[31].

**Acknowledgements**

The authors are grateful to J. Stephen, A Hyndman, G. Williams, and P. Murmu for technological help for SQUID and Hall Effect measurements. This research is supported by the New Economy Research Fund (contract VICX0808) and the Marsden Fund (08-VUW-030).

**Author contributions**

F.N., B. J. R., H. J. T. proposed the present study and organized the research project. The AlN templates on Si(111) substrates were made by using the MBE method at CRHEA/CNRS laboratory by F.N. and F.S., while the growth of GdN films and AlN capping layer were carried out at Victoria University of Wellington by F.N. F.N., S.S. and L.H. worked on nuclear analysis data acquisition at CNEBG/IN2P3, and analysed the spectra. F.N., N.O.V.P, B.J.R. and H.J.T. carried out the electrical transport and magnetic data, analysed and interpreted the results. F.N., B. J. R., H. J. T. wrote the manuscript. All authors discussed the results and commented on the manuscript.

**Competing Financial Interests Statement**

The authors declare no competing financial interests.

**Figure Legends**

Figure 1 **Real time investigation by reflection high-energy electron diffraction of the epitaxial growth of GdN.** **a**, Schematic view of thin-film specimens of GdN studied in this letter. **b**, Variation of the in-plane surface lattice parameter during the growth of GdN on AlN as measured by RHEED. The inset shows the RHEED pattern along the Si [110] azimuth of a 14 nm thick GdN layer.

Figure 2 **X-ray diffraction scan of an AlN-capped GdN film grown on an AlN/Si(111) sequence.** This typical $\Theta$-$2\Theta$-scan XRD scan shows that in addition to the peaks of AlN and Si only the (111) and (222) reflections of GdN are present in these films. These data demonstrate the wurtzite (0001) surface of the AlN buffer layer favours the growth of fully (111)-oriented GdN film.

Figure 3 **Rutherford backscattering spectroscopy measurements in random and channelling conditions of AlN/20.5 nm GdN/AlN/Si(111) sequence**. The random (blue) and aligned (black) RBS spectra are recorded using a 2 MeV $^4$He$^+$ ion beam with 160° scattering angle. The random spectrum was done with a rotating sample with 5° tilt angle. The lines (labelled Gd, Al, Si and N) indicate the onset of backscattering energy for these elements at the surface. The channelling is highlighted by the very similar reduction of both Gd, Si substrate and the AlN buffer layer signals, while as expected there is no evidence of channelling in the polycrystalline AlN cap layer.

Figure 4 **In-plane magnetic properties of a 14 nm thick GdN layer. a**, The temperature-dependent magnetization and inverse susceptibility in an applied field of 25 mT. **b**, The field-dependent magnetisation at 10K. The inset is a magnification at low field showing a coercive field of about 125Oe.

Figure 5 **Resistivity of a 14 nm GdN film.** The temperature dependence of the resistivity shows a negative temperature coefficient broken by the well-established anomaly at the Curie temperature.

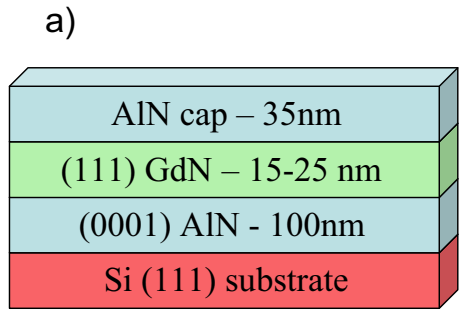
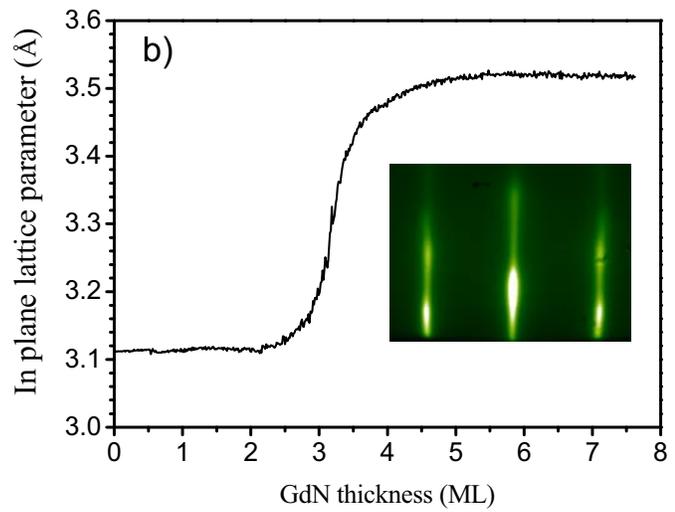

Figure 1, F. Natali *et al.*

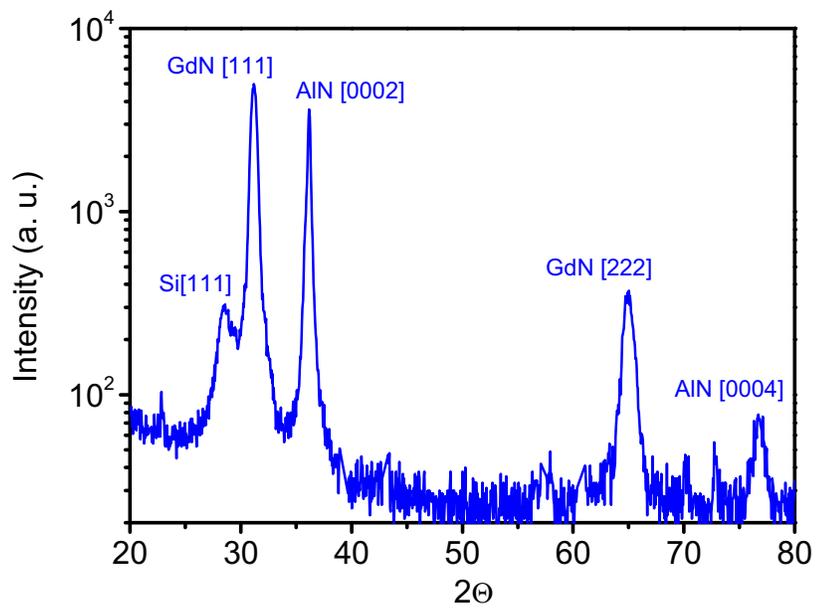

Figure 2, F. Natali *et al.*

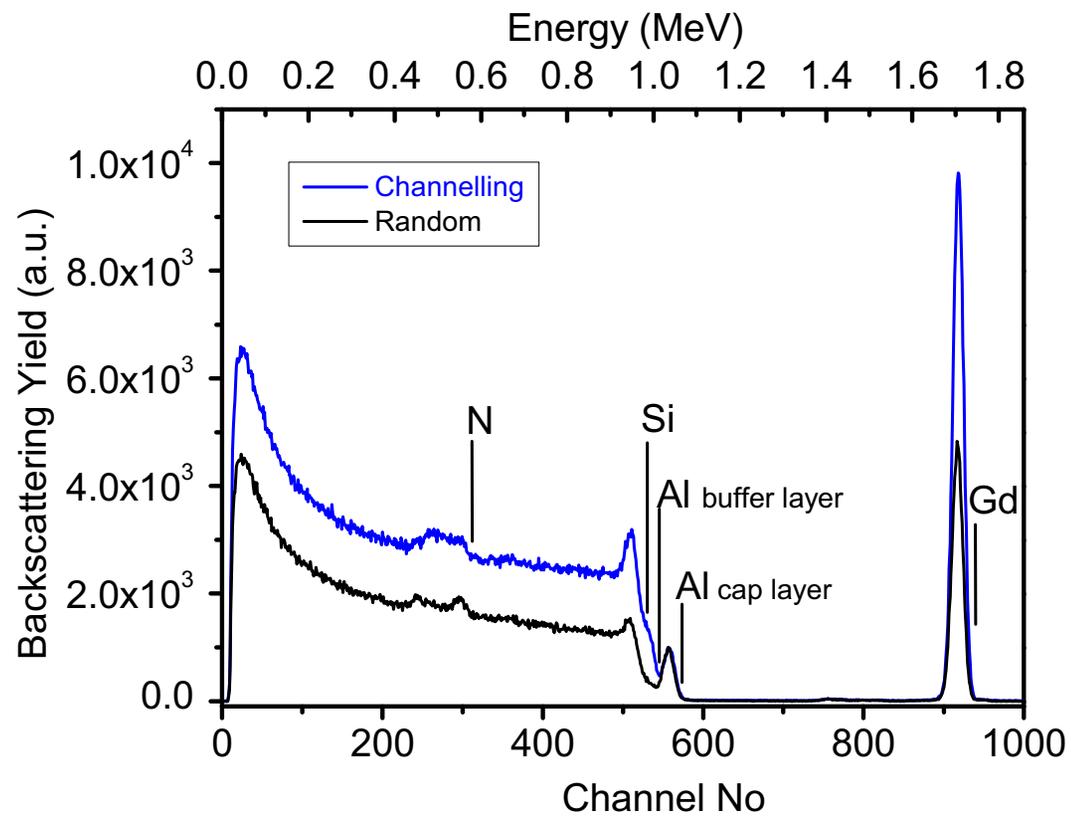

Figure 3, F. Natali *et al.*

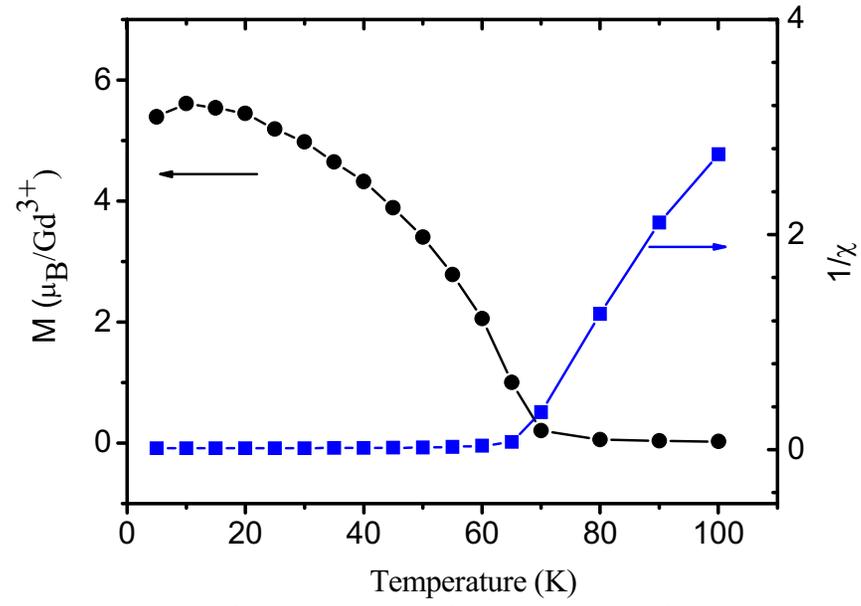
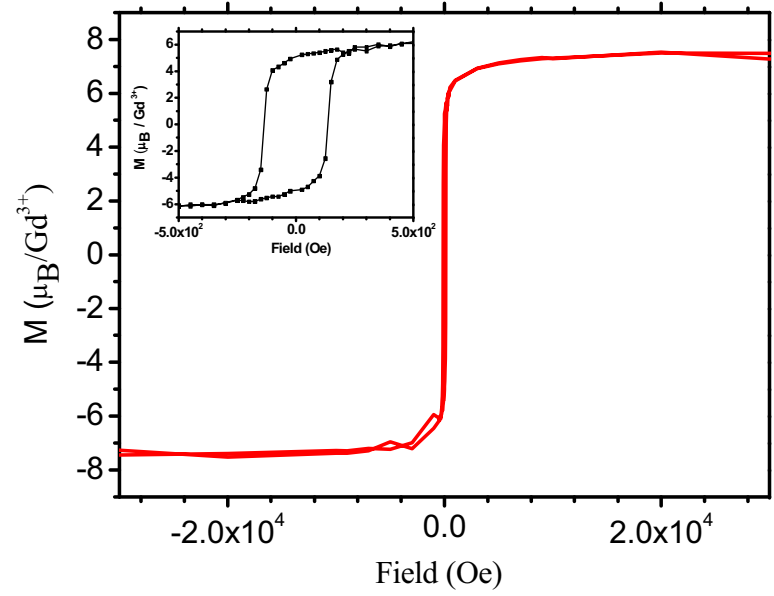

Figure 4, F. Natali *et al.*

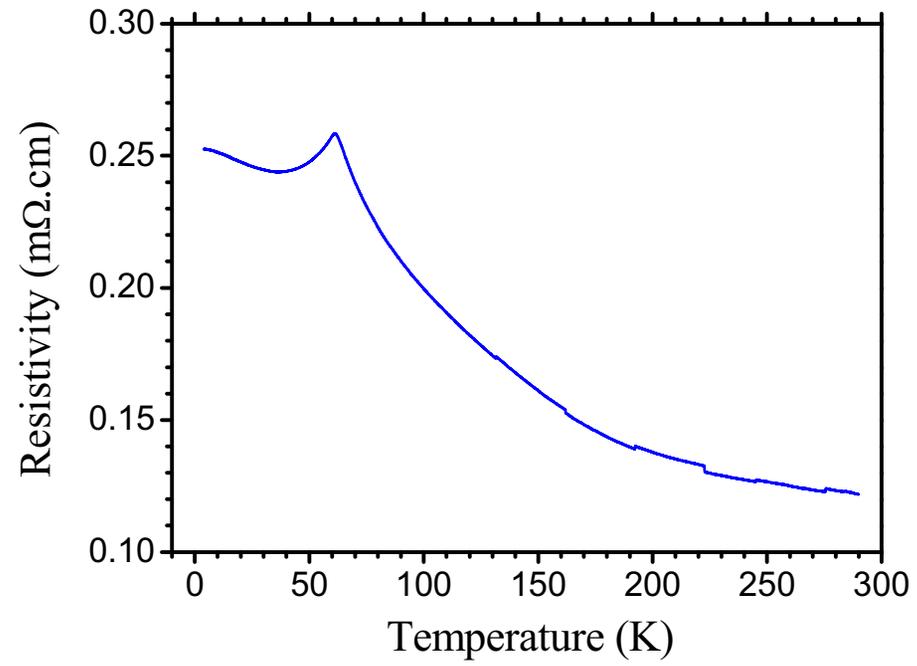

Figure 5, F. Natali *et al.*